\documentclass[a4paper,11pt]{article}
\usepackage[a4paper,top=1in,bottom=1in,left=1in,right=1in]{geometry}
\usepackage{amsthm,amssymb,amsmath,latexsym,amsfonts}

\begin{document}

\clubpenalty=10000
\widowpenalty = 10000

\hyphenation{al-go-rithm coef-fi-cient comple-xi-ties con-tem-po-rary
dis-crimi-nant Fel-low-ship im-ple-men-ta-tion
Ma-no-cha Min-kow-ski multi-graded New-ton opti-mal
pan-a-cea po-ly-no-mial po-ly-no-mials po-ly-topes Pres-i-den-tial
ran-do-mi-za-tion re-sul-tant sys-tem vo-lume} 

\newtheorem{theorem}{Theorem}[section]
\newtheorem{lemma}[theorem]{Lemma}
\newtheorem{corollary}[theorem]{Corollary}
\newtheorem{definition}[theorem]{Definition} 
\newcommand{\CC}{\mathbb{C}} \newcommand{\PP}{\mathbb{P}} 
\newcommand{\RR}{\mathbb{R}} \newcommand{\ZZ}{\mathbb{Z}}
\newcommand{\MV}{{\mbox{\it\em MV}} }
\newcommand{\vol}{{\mbox{\it\em vol}} }

\title{Compact Formulae in Sparse Elimination\\
{\small [Abstract of Invited talk at ACM ISSAC, July 2016]}}

\author{
Ioannis Z.~Emiris\footnote{Member of 
AROMATH, a joint team between INRIA Sophia-Antipolis, France,
and NKU Athens.
Partially supported by H2020 Marie Sk\l{}odowska Curie
European Training Network ARCADES (Algebraic Representations for
Computer-Aided Design of complEx Shapes), 2016-2019, project number: 675789.
}\\
{National and Kapodistrian University of Athens, and Athena Research Center, Greece } \\
{\tt emiris@di.uoa.gr}
}

\maketitle
\begin{abstract}
It has by now become a standard approach to use the
theory of sparse (or toric) elimination,
based on the Newton polytope of a polynomial,
in order to reveal and exploit the structure of algebraic systems.
This talk surveys compact formulae, including
older and recent results, in sparse elimination.

We start with root bounds
and juxtapose two recent formulae: a
generating function of the m-B\'ezout bound and
a closed-form expression for the mixed volume by means of a matrix permanent.  
For the sparse resultant, a bevy of results have established determinantal
or rational formulae for a large class of systems, starting with Macaulay.
The discriminant is closely related to the resultant but admits no compact formula except for very simple cases. We offer a new determinantal formula for the discriminant of a sparse multilinear system arising in computing Nash equilibria.
We introduce an alternative notion of compact formula, namely the Newton polytope of the unknown polynomial. It is possible to compute it efficiently for sparse resultants, discriminants, as well as the implicit equation of a parameterized variety. 
This leads us to consider implicit matrix representations of geometric objects.

\medskip

\end{abstract}

\section{Introduction}

It is today a standard approach to use the theory of sparse (or toric) elimination, based on the Newton polytope of a polynomial, in order to reveal and exploit the structure of algebraic systems. 
Consider Laurent polynomials in $N$ variables:
$$
f_i = \sum_{j=1}^{k_i}{ c_{ij} {x}^{a_{ij}}}, \quad c_{ij}\ne 0,
$$
where $x=(x_1,\dots,x_N)$, $x^e =\prod_i x_i^{e_i}$, and
$\{ a_{i1}, \dots, a_{ik_i} \} \subset \ZZ^N$
is the {\em support} of $f_i$.
Let $Q_i$ be the {\em Newton polytope} of $f_i$ defined
as the convex hull of the support.

A more constrained form of structure corresponds to
multihomogeneous systems, which are defined on a product
\begin{equation} \label{eq:pps}
\PP^{n_1}\times\PP^{n_2}\times\cdots\times\PP^{n_S}
\end{equation}
of projective spaces: the variables are partitioned into $S$ subsets,
or blocks, so that each equation is homogeneous 
in each block of $n_j+1$ homogeneous variables.
In contrast, semi-mixed systems are those whose polynomials
can be partitioned into subsets, or blocks, with fixed monomial support.
In sparse elimination, each subset of polynomials
has the same Newton polytope.
In the case of semi-mixed multihomogeneous systems, 
the polynomials in each block are homogeneous of the same degree 
per block of variables. 

An interesting example of a semi-mixed multihomogeneous system appears
in game theory and was studied in \cite{EmiVid14,McMc97}.
Consider a game of $S$ players, each with $m_i$ options or strategies.
The $j$-th player plays a {totally mixed strategy} when the
corresponding strategy is chosen randomly with nonzero probabilities
$p^{(j)}_1,\ldots,p^{(j)}_{m_j}.$
A totally mixed Nash equilibrium (TMNE) 
is a combination of such strategies so that no player can improve
his payoff by unilaterally choosing other (pure or mixed) strategy.  
We express payoff $P_j$ of the $j$-th player choosing the $i$-th
strategy by the polynomial
\begin{equation} \label{eq:tmne}
\sum_{k_1,\ldots,k_{j-1},k_{j+1},\dots,k_S}
	a^{(j)}_{\ldots,k_{j-1},i,k_{j+1},\dots}
p^{(1)}_{k_1}\cdots p^{(j-1)}_{k_{j-1}}
	p^{(j+1)}_{k_{j+1}} \cdots p^{(S)}_{k_S} .
\end{equation}
Here $a^{(j)}_{k_1,k_2,\ldots,k_S}$ denotes the predefined payoff
of the player, assuming each player chooses the pure option 
$k_{\ell} \in \{1,\ldots,m_{\ell}\}$. 
The equations imply that the payoff $P_j$ does not depend on own strategy
$i$, as long as other players do not change their strategies.
Eliminating the $P_j$'s leads to a multilinear system with $n_j=m_j-1$
in space~(\ref{eq:pps}).
Actual probabilities are determined from the normalizing conditions
$ \sum_{i=1}^{m_j} p^{(j)}_i = 1$, for $j\in\{1,\dots,S\}.$

\section{Root bounds}

Bounds on the number of common roots of an algebraic system
admit different types of compact formulae.  
We start with classic bounds named after B\'ezout, 
tight for (multi)homogeneous systems.
The most general such bound is the following m-B\'ezout bound,
stated for dehomogenized system. 

\begin{theorem}[m-B\'ezout bound]\label{thm:mBez}
Consider a system of $N$ equations in $N$ affine variables,
partitioned into $S$ subsets so that the $j$-th subset includes
$n_j$ affine variables, and $N=n_1+\cdots +n_S$.
Let $d_{ij}$ be the degree of the $i$-th equation in
the $j$-th variable subset, for $i=1,\dots,N$ and $j=1,\dots,S$.
If the system has finitely many complex roots in space~(\ref{eq:pps}), 
the coefficient of $x_1^{n_1}\cdots x_S^{n_S}$ in
\begin{equation} \label{eq:mhbprod}
\prod_{i=1}^N (d_{i1}x_1 + \cdots + d_{iS}x_S ) 
\end{equation}
bounds the number of roots.
For generic coefficients this bound is tight.
\end{theorem}

We shall give a generating function for the m-B\'ezout number by
extending results in \cite{McMc97,McLn99,VidunasDerang}.
MacMahon's Master theorem \cite{McMah16} is a powerful combinatorial result.
\begin{theorem}[MacMahon's Master Theorem] {\em\cite{McMah16}}
Let $A=(a_{ij})$ be a complex $S\times S$ matrix,
$x_1,\ldots,x_S$ be formal variables, and $V$
denote the diagonal matrix with nonzero entries $x_1,\ldots,x_S$.
The coefficient of $x_1^{n_1}\cdots x_S^{n_S}$ in
\begin{equation} \label{eq:mhms0}
\prod_{j=1}^{S} (a^{}_{j1}x^{}_1+\cdots+a^{}_{jS}x^{}_S)^{n_j}
\end{equation}
equals the coefficient of $x_1^{n_1}\cdots x_S^{n_S}$ in
the multivariate Taylor expansion of
\begin{equation} \label{eq:mhms}
f(x_1,\ldots,x_S)=\frac{1}{\det (I-VA)}
\end{equation}
around $(x_1,\ldots,x_S)=(0,0,\ldots,0)$.
\end{theorem}

Let us consider semi-mixed multihomogeneous systems, where
the number of equation blocks equals
the number $S$ of variable blocks, and we have a varying number of 
equations per block; more importantly, the degree per block is arbitrary.
Generically, this system has a finite number of solutions.
MacMahon's Theorem 
yields a multivariate generating function for the m-B\'ezout bound
of these systems,
which could be helpful when one seeks root counts for a family of systems.  

\begin{theorem} \label{tm:smmh} {\em\cite{EmiVid14}}
Consider a multihomogeneous system on space
(\ref{eq:pps}) of $N=n_1+ \cdots +n_S$ equations, where
the equations are partitioned into $S$ subsets 
of exactly $n_1,\ldots,n_S$ equations.  
We assume that for any $i,j\in\{1,2,\ldots,S\}$, the polynomial equations 
in the $i$-th subset have degree $a_{ij}$ in the variables of the $j$-th variable subset.
Let $A$ be the $S\times S$ matrix defined by the $a_{ij}$'s.
Then the m-B\'ezout bound for the multihomogeneous system equals the
coefficient of $x_1^{n_1}\cdots x_S^{n_S}$
in the multivariate Taylor expansion of
$$
{1} / {\det (I-V A)}
$$
around $(x_1,\ldots,x_S)=(0,0,\ldots,0)$.
\end{theorem}

For TMNE system~(\ref{eq:tmne}),
we have $a_{ii}=0$, and $a_{ij}=1$ for $i\neq j$.
Then 
\[
I-VA= \left[ \begin{array}{ccccc}
1 & -x_1 & -x_1 & \cdots & -x_1 \\
-x_2 & 1 & -x_2 & \cdots & -x_2 \\
\vdots & \vdots & \ddots & \ddots & \vdots \\
-x_S  & -x_S & \cdots & -x_S & 1
\end{array} \right] ,
\]
and
$$
{1} / {\det (I-V A)} =
\frac{1}{1-\sigma_2-2\sigma_3-\cdots-(S-1)\sigma_S},
$$
where $\sigma_j$ (for $j=2,\ldots,S$) denotes the $j$-th
elementary symmetric polynomial in $x_1,\ldots,x_S$.

Now we pass to the most general root bound in sparse elimination theory,
namely mixed volume: there are efficient algorithms but,
in certain cases, the mixed volume computation is reduced to a
matrix permanent.
The toric root bound is named after
Bernstein, Khovanskii, and Kushnirenko (BKK) \cite{Bern75}:

\begin{theorem} \label{th:bkk}
For $f_1, \dots f_N \in \CC[ x_1^{\pm 1}, \dots, x_N^{\pm 1}]$
with Newton polytopes
$Q_1, \dots, Q_N$, the number of common isolated solutions,
multiplicities counted, in the corresponding toric variety,
which contains $( \CC^{*})^N$ as a dense subset,
does not exceed $\MV( Q_1, \ldots$, $Q_N)$,
independently of the toric variety's dimension.
\end{theorem} 

It is known that mixed volume is related to the permanent.
In fact, by reduction to the latter one shows that mixed volume is
$\#$P-complete.
For a multihomogeneous system on space~(\ref{eq:pps}) of
$N=n_1+\cdots+n_S$ equations,
assume the $i$-th equation has degree $a_{ij}$
in the $j$-th variable block.
Let $A=(a_{ij})$ be an $N\times N$ matrix 
with the columns repeated $n_j$ times for each $j\in\{1,2,\ldots,S\}$.
Then the mixed volume equals \cite[Thm~2]{McLn99}
$$
\frac{1}{n_1!\cdots n_S!} \; \mbox{perm } A.
$$ 
A generalization to arbitrary (non-homogeneous) systems follows.
For each block of variables there is a $n_j$-dimensional polytope
$\Gamma_j$, each in a separate complementary space, $1\le j\le S$.
Assume the Newton polytopes are direct products
of scalar multiples (by $a_{ij}$) of the $\Gamma_j$'s.
We obtain an algebraic system on the product of toric 
varieties corresponding to $\Gamma_j$, of dimension $n_j$.

\begin{theorem} {\em \cite{EmiVid14}}
Let $A$ be the matrix of $a_{ij}$'s with columns repeated $n_j$ times.
If there are $N=\sum_{j=1}^S n_j$ equations, then
$$
\MV(Q_1,\dots,Q_N) = \mbox{\em perm}\, A \, \prod_{j=1}^S \vol(\Gamma_j)
,\quad Q_i=\prod_{j=1}^S a_{ij}\Gamma_j.
$$
\end{theorem}

\section{Resultant formulae}

For the sparse resultant, a bevy of results have established matrix and,
more particularly, determinantal formulae for a large class of systems.
An introductory survey can be found in \cite{MoroShak10}.

Resultants provide efficient ways for studying and solving polynomial
systems by means of matrices
whose determinant is a non-trivial multiple of the resultant.  
They are most efficiently
expressed by a generically non-singular matrix, whose determinant is
the resultant polynomial or, when this is impossible, by also
specifying a minor $M'$ of resultant matrix $M$
which divides the determinant of the latter so as to yield a
rational formula for the resultant $R$:
$$
R =\, \det M\, /\, \det M' .
$$
Macaulay's classical result \cite{Mac1902}
establishes such rational formula for arbitrary dense systems, and 
D'Andrea, 100 years later \cite{DAnd02},
for arbitrary systems in the context of sparse elimination.
His recursive construction has been simplified
in certain cases \cite{EmiKonJsc}.

For two univariate polynomials there are matrix formulae
named after Sylvester and B\'ezout, whose determinant is equal to the
resultant; we refer to them as determinantal formulae.
The largest family admitting determinantal formulae is a class of
multihomogeneous systems studied in \cite{StZe,WeZe},
see also~\cite[Sect.13.2]{GKZ}, where, each block of variables 
contains a single dehomogenized variable or is of linear total degree.
These systems are moreover unmixed, in other words all polynomials have the
same Newton polytope; they are called multigraded.

The multigraded resultant matrices were made explicit in
\cite{DicEmi03,EmiMan09}, which offered determinantal formulae of 
Sylvester, B\'ezout and hybrid types, the latter containing
blocks of both Sylvester and B\'ezout types.  
In \cite{DicEmi03}, it is 
shown that there exists a determinantal pure B\'ezout-type resultant
formula if and only if there exists such a Sylvester-type formula.
The B\'ezout-type matrices generalize those identified in \cite{ChtKap00}.
In \cite{EmiMan09} was proven the existence of determinantal formulae
for systems whose Newton polytopes are scaled copies of one polytope.  

In \cite{DADi01}, hybrid formulae were proposed in the mixed homogeneous case.
The most complete study of hybrid formulae is \cite{EisSch03} where
resultants are given as a matrix determinant or a Pfaffian, i.e.\
a determinant square root, 
for unmixed systems of 3 dense polynomials, 
for up to 5 polynomials of degree up to 4, 6, or 8,
and for~6 quadratic polynomials.

Some studies have focused on three polynomials.
In the unmixed case a determinantal formula is established in 
\cite{Khet03}, more direct than \cite{EisSch03} and
generalizing the Macaulay-style formula 
obtained for the dense case in~\cite{Joua97}, where
the numerator matrix has one row corresponding to the affine Jacobian.
For Newton polygons which are scaled copies of a single one,
the smallest Sylvester-type matrices are obtained by using a row
containing the coefficients of the toric Jacobian~\cite{DAnEmi02}.  
In \cite{ZhGo00} they construct determinantal Sylvester-type formulae for
unmixed systems whose Newton polygon is a rectangle from which
smaller rectangles have been removed at the corners.

\section{Discriminant formulae}

Discriminants are crucial in studying well-constrained algebraic systems,
the system's zero set, its singularities,
and other properties of the system.
The discriminant is closely related to the resultant but admits
no compact formula except for very simple cases.
Below we offer a new determinantal formula for a specific system.
Recent work has shed light on the degree of mixed discriminants
\cite{CCDRS,DiEmKa} as a function of the Newton polytopes of
the corresponding polynomials.
An important open question is to obtain an explicit formula
for multilinear systems \cite{SturmfHur}.

In the context of sparse elimination the discriminant is defined for
a set of polynomials $F_i$ with fixed supports $A_i$, hence it is 
denoted $\Delta_{A_1, \dots, A_n}(F_{1}, \dots ,F_{n})$
or $\Delta(F_{1}, \dots ,F_{n})$.

\begin{definition} 
The \textit{mixed discriminant} $\Delta(F_{1}, \dots ,F_{n})$
of $n$ polynomials in $n$ variables with fixed supports
$A_1, \dots, A_n$ in $\mathbb{Z}^n$ is the irreducible polynomial
(with integer coprime coefficients, defined up to sign)
in the coefficients of the $F_i$ which
vanishes whenever the system $F_1=\cdots=F_n=0$ has a multiple root
(i.e., a root which is not simple) with nonzero coordinates,
provided his discriminantal variety is a hypersurface;
otherwise, the mixed discriminant is equal to the constant $1$.
\end{definition}
Clearly, the zero locus of the mixed discriminant is the variety of ill-posed systems,
i.e., systems with at least one multiple root. 

We focus on systems over $\PP^1\times\PP^1\times\PP^1$ where each equation
is missing one variable, motivated by the study of TMNE in games
of~3 players with~2 pure strategies each, see polynomials~(\ref{eq:tmne}).
The bilinear system is as follows:
\begin{align}
F_1: && a_0y_1z_1+a_1y_1z_0+a_2y_0z_1+a_4y_0x_0 \, = & \, 0, \nonumber \\
F_2: && b_0x_1z_1+b_1\,x_1z_0+b_3\,x_0z_1+b_4x_0z_0 \, = & \, 0,\nonumber\\
F_3: &&  c_0\,x_1y_1+c_2\,x_1y_0+c_3\,x_0y_1+c_4x_0y_0 = & \, 0 .\nonumber
\end{align}
The projective coordinates $(x_1:x_0)$, $(y_1:y_0)$, $(z_1:z_0)$ represent 
the unknown probability ratios for the two options of each player;
the coefficients $a_i$ equal the differences in the given payoffs
(between the two options) of the first player, for the 4 combinations
of pure choices of the other two players;
similarly, $b_i$, $c_i$ equal the differences in the given payoffs
(between their two options) of the other two players.

The following theorem gives a $6\times 6$ determinantal expression for the discriminant.
The formula is of Sylvester-type because the matrix entries are
either zero or equal to a polynomial coefficient.  

\begin{theorem} {\em\cite{EmiVidDisc}}
The discriminant $\Delta(F_1,F_2,F_3)$ equals 
$$
\det  \left[ \begin{array}{cccccc}
0 & 0 & c_0 & c_1 & b_0 & b_1 \\
0 & 0 & c_2 & c_4 & b_3 & b_4 \\
c_0 & c_2 & 0 & 0 & a_0 & a_2 \\
c_1 & c_4 & 0 & 0 & a_3 & a_4 \\
b_0 & b_3 & a_0 & a_3 & 0 & 0 \\
b_1 & b_4 & a_2 & a_4 & 0 & 0
\end{array} \right].
$$
\end{theorem}

\section{Sparse implicitization}

We argue that in sparse elimination it makes sense to consider
an alternative notion of compact formula, namely
the Newton polytope of the unknown polynomial.
It typically allows computation of the polynomial by interpolation of the
coefficients.
It is possible to compute it efficiently for
sparse resultants~\cite{EmFiKoPeJ}, discriminants, 
as well as the implicit equation of a parameterized variety. 

This section focuses on the latter case.
Given (a superset of) the Newton polytope of the implicit equation,
one constructs an interpolation matrix, which allows for various
geometric operations to be efficiently computed on this matrix without
need of developing the monomial representation of the implicit equation.
The representation by interpolation matrices has been developed
for parametric plane curves, surfaces, and hypersurfaces in
the context of sparse elimination \cite{EmKaKoGmod,EmKaKoLBspm}.
It generalizes to space curves and, generally,
objects of codimension higher than~1 by means of the Chow form,
which is a generalization of resultants to $N$-variate systems
with more than $N+1$ polynomials \cite{EmiGavKon}.

Another powerful matrix representation of geometric objects
is based on syzygies.
The corresponding theory, including $\mu$-bases, has been developed for
parametric models \cite{BusLBa10,BuCoDA,SedChe95}.
The computation of syzygies can exploit the
sparse elimination setting \cite{BotDic16} and,
moreover, can be achieved for point cloud models \cite{EmiGavKon}.  
Roughly, syzygies of a sufficiently high degree define
moving curves or surfaces that ``follow" the parametric object. 
One forms the matrix expressing these moving curves and/or surfaces,
which represents the geometric object
because 
its rank at some point $p$ drops if and only if $p$ belongs to
the algebraic closure of the image of 
the parameterization \cite{BusLBa10}.


\end{document}